\documentclass[online]{vgtc}
\usepackage[utf8]{inputenc}
\usepackage[a4paper, total={18cm, 26cm}]{geometry}
\usepackage{graphicx}
\graphicspath{ {./figures/} }
\usepackage{subfig}
\usepackage{hyperref}
\usepackage[svgnames]{xcolor}
\usepackage[most]{tcolorbox}
\usepackage{geometry}
\usepackage{enumitem}

\usepackage{microtype}                 
\PassOptionsToPackage{warn}{textcomp}  
\usepackage{textcomp}                  
\usepackage{mathptmx}                  
\usepackage{times}                     
\usepackage{cite}                      
\usepackage{tabu}                      
\usepackage{booktabs}                  
\usepackage{enumitem}
\usepackage{mathpazo}

\newcommand{\subscript}[2]{$#1 _ #2$}
\newcommand{\dg}[1]{\subscript{DG}{#1}}
\newenvironment{designgoals}%
  {\begin{enumerate}[label=\subscript{DG}{{\arabic*}}]%
    \setlength{\itemsep}{1.5pt}%
    \setlength{\parskip}{2pt}}%
  {\end{enumerate}}

\vgtcinsertpkg

\title{Corvo: Visualizing CellxGene Single-Cell Datasets in Virtual Reality}

\author{
Luke Hyman\thanks{e-mail: lukejhyman@gmail.com}\\
    \parbox{1.5in}{\scriptsize \centering 
    Chan Zuckerberg Biohub \\
    Center for Systems Biology Dresden}
\and        
Ivo~F.~Sbalzarini\\
    \parbox{1.5in}{\scriptsize \centering Technische Universit\"at Dresden \\ Center for Systems Biology Dresden \\ MPI-CBG, Dresden }
\and        
Stephen Quake\\
    \parbox{1.5in}{\scriptsize \centering Dept. of Bioengineering, Stanford University \\ Chan Zuckerberg Biohub }
\and 
Ulrik~G\"{u}nther\thanks{e-mail: ulrik.guenther@hzdr.de}\\ 
    \parbox{1.5in}{\scriptsize \centering CASUS, G\"orlitz\\ Center for Systems Biology Dresden \\ MPI-CBG, Dresden  }
}

\abstract{
The CellxGene project has enabled access to single-cell data in the scientific community, providing tools for browsed-based no-code analysis of more than 500 annotated datasets. However, single-cell data requires dimensional reduction to visualize, and 2D embedding does not take full advantage of three-dimensional human spatial understanding and cognition. Compared to a 2D visualization that could potentially hide gene expression patterns, 3D Virtual Reality may enable researchers to make better use of the information contained within the datasets. For this purpose, we present \emph{Corvo}, a fully free and open-source software tool that takes the visualization and analysis of CellxGene single-cell datasets to 3D Virtual Reality. Similar to CellxGene, Corvo takes a no-code approach for the end user, but also offers multimodal user input to facilitate fast navigation and analysis, and is interoperable with the existing Python data science ecosystem. In this paper, we explain the design goals of Corvo, detail its approach to the Virtual Reality visualization and analysis of single-cell data, and briefly discuss limitations and future extensions.}
\begin{document}

\maketitle

\section{Introduction}
The growing availability of and interest in single-cell transcriptomics data has spurred the development of resources and tools to expand its access and ease of analysis. One such tool is the Chan-Zuckerberg \emph{Cell by Gene Discover (CellxGene)} online platform for dataset aggregation and no-code analysis \cite{ambrosecarrCellxgenePerformantScalable2021}. CellxGene offers an open-access database of currently more than 500 fully annotated single-cell datasets, with browser-based analysis tools that replicate most of the commonly-used workflows from libraries such as Scanpy \cite{wolfSCANPYLargescaleSinglecell2018} or Seurat \cite{butlerIntegratingSinglecellTranscriptomic2018}.

The datasets from CellxGene contain the transcriptome of hundreds of thousands of individual cells, with on the order of 10s of thousands of genes per cell, and represents them through dimensional reduction techniques like UMAP \cite{mcinnesUMAPUniformManifold2020} and tSNE \cite{vanderMaaten:2008Visualizing}. These represent higher dimensional information in a new 2D/3D coordinate system, where cells with similar properties in higher dimensions are clustered together. Dimensional reduction to 2D has been the most commonly-used version, as 2D plots can easily be presented and navigated on a regular screen. However by going to 3D, adding an additional dimension, more higher dimensional structure has been shown to lift out of the plot, potentially allowing for gene expression patterns not previously visible to be discovered\cite{shamitsonejiCellexalVRVirtualReality2021}. For 3D data in turn, visualization and interaction in Virtual Reality (VR) has been shown to help with spatial understanding and cognition\cite{Slater:2016552}.


 Here, we present \emph{Corvo}, a Java Virtual Machine (JVM)-based visualization and analysis tool for dimensionally-reduced single cell data. We have designed Corvo with the following four design goals in mind:
 \begin{designgoals}
     \item \emph{Interoperability} -- with data science workflows usually at home in the Python ecosystem, Corvo should be able to interface with Python easily.
     \item \emph{Ease-of-use} -- importing any CellxGene dataset should not require manual user intervention or coding.
     \item \emph{Multimodality} -- while VR can ease spatial understading, other tasks, like textual input, are rather complicated. Corvo should therefore be able to use e.g. voice input for gene name search. 
     \item \emph{Versatility} -- users should be able to access analysis tools and statistical tests easily.
\end{designgoals}

\begin{figure}[ht!] 
    \centering
    \includegraphics[width=\linewidth]{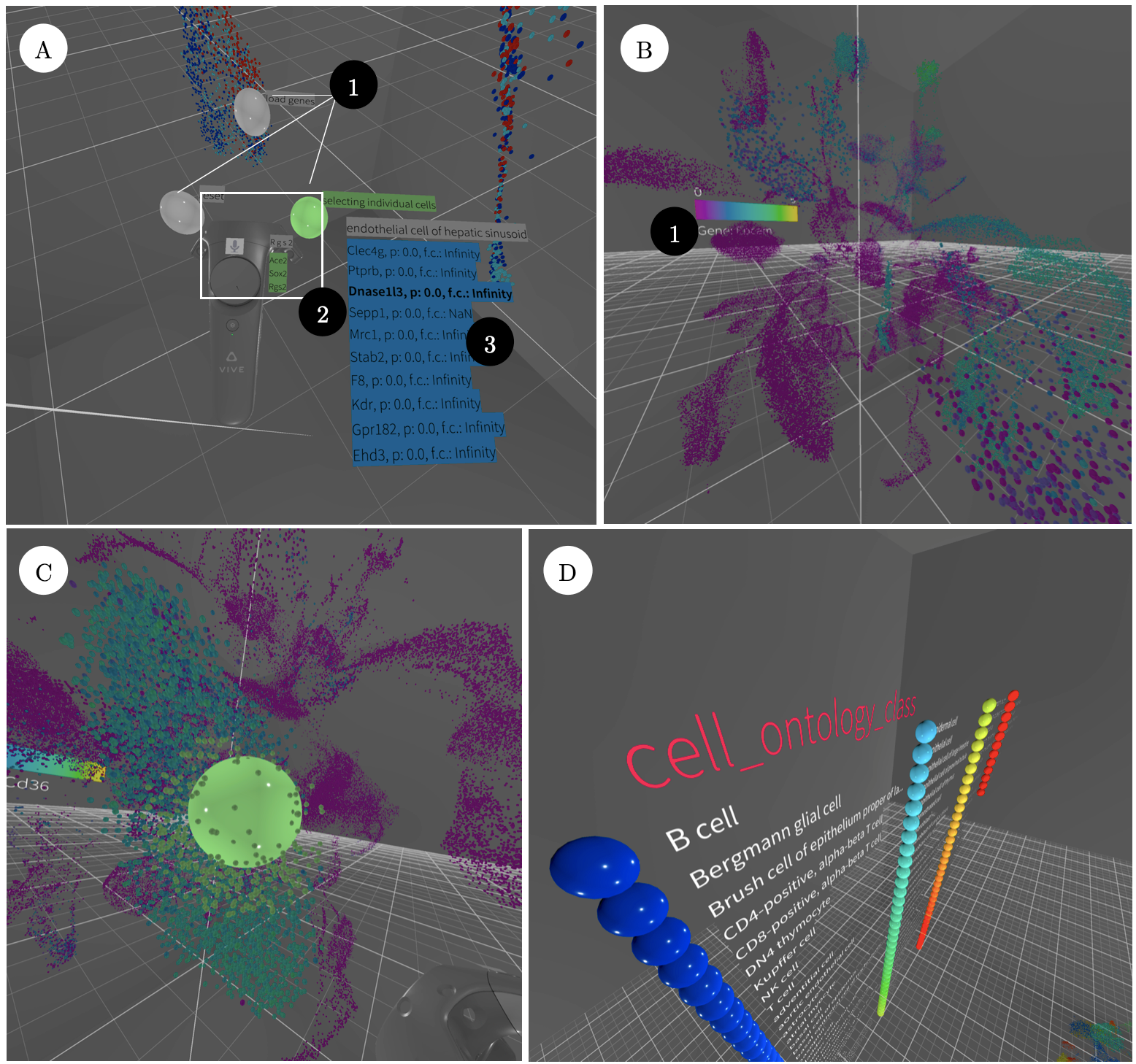}
    
    \caption{Examples of Corvo's VR interface: \textbf{A}: menu (1), press-to-talk voice interface, with genes recognized from voice input listed (2), list of precomputed genes (3) \textbf{B}: UMAP plot with gene expression scale (1) \textbf{C}: interaction / selection sphere \textbf{D}: annotation key. See Section \ref{sec:VRinterface} for a detailed description.\label{fig:VRinterface}}
\end{figure}
\section{Related Work}
A number of projects have already endeavoured to build VR single-cell data analysis tools, allowing the 3D structures to actually be viewed in stereoscopic 3D. Yet none fulfil our design goals. Two such projects, CellexalVR \cite{shamitsonejiCellexalVRVirtualReality2021} and Theia \cite{gregoryj.hannonExplorationAnalysisMolecularly2021}, use dedicated VR hardware to facilitate a single-cell data analysis workflow entirely from within Virtual Reality. However, the need for users perform manual steps for dataset preprocessing (in either Python or R) adds a similar barrier to entry that CellxGene sought to solve in the 2D space, violating \dg{1} and \dg{2}. Starmap \cite{joshuaw.k.hoStarmapImmersiveVisualisation2018} and singlecellVR \cite{lucapinelloSinglecellVRInteractiveVisualization2021} both use inexpensive smartphone-based head mounted displays for VR, which limits user interaction and on-the-fly data analysis due to hardware limitations. They can therefore not entirely replicate a normal single-cell data workflow, and so do not fulfil \dg{4}. The two projects targeting dedicated VR headsets -- CellexalVR \cite{shamitsonejiCellexalVRVirtualReality2021}, and Theia \cite{gregoryj.hannonExplorationAnalysisMolecularly2021}) --  have implemented a broader set of analytical features. However, they require additional manual conversion steps of the user-supplied datasets, again violating \dg{2}. While starmap offers voice input, it is limited to starting and stopping animations. None of the other mentioned software packages offer multimodal user input (\dg{3}). In contrast to Corvo's current implementation though, they are agnostic to the dataset source, while Corvo focuses on CellxGene datasets. 

\section{Corvo}
\begin{figure}
    \centering
    \includegraphics[width=\linewidth]{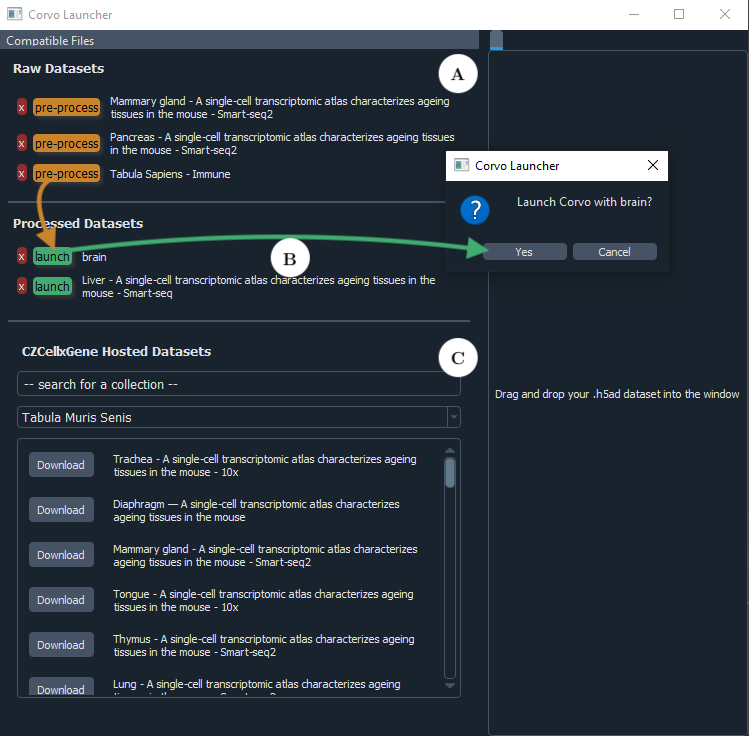}%
    \caption{The Corvo launcher interface: \textbf{A}: List of downloaded CellxGene datasets not yet processed \textbf{B}: list of processed datasets ready for launch \textbf{C}: dataset collection search box and dropdown list, with list of contained datasets for download. See Section \ref{sec:Datasets} for a detailed description.\label{fig:LauncherInterface}}
\end{figure}


Corvo is built using the scenery rendering framework \cite{guntherSceneryFlexibleVirtual2019} and written in Kotlin\footnote{See \href{https://kotlinlang.org}{kotlinlang.org}}. The core components of Corvo are packaged into a single JAR for easy distribution, and we provide a Python-based front-end for easy interoperability with existing analysis packages (\dg{1}). Data is read in directly from the processed \texttt{h5ad} files by Corvo using the jHDF5 library\footnote{See \href{https://sissource.ethz.ch/sispub/jhdf5}{sissource.ethz.ch/sispub/jhdf5}}. This has the advantage of being faster and more memory efficient than converting to text-based formats, such as JSON or CSV, as other solutions require. Gene expression data is also kept in its native sparse format and reconstructed on demand to save memory. 

\subsection{Dataset Management}
\label{sec:Datasets}
For a completely code-free usage, a PyQt5-based user interface is used to launch the VR experience with a selected dataset. In the launcher, Corvo automatically acquires a list of the available CellxGene datasets and presents them for the user to download. Once downloaded and optimised for Corvo, the optimised representation\footnote{CellxGene datasets don't always adhere to the same layout. We preprocess them automatically into a unified format to make downstream analysis easier and faster.} is stored for repeated use, after which the raw datasets may safely be deleted from within the launcher. The Corvo launcher is shown in Figure \ref{fig:LauncherInterface}. From there, any dataset can be selected, and the VR experience launched with it.

\subsection{User Interface and Data Analysis}
\label{sec:VRinterface}
In the VR scene, cells are shown at the locations corresponding to the 3D coordinates originating from dimensional reduction. They are rendered using instancing, allowing Corvo to render the data of about 500000 cells while maintaining VR-compatible framerates on commodity hardware.
The 3D plot is placed centrally in the scene and can be resized, rotated, or translated for walking exploration at human scale, or by flying through with the joysticks of the VR headset's controller (Figure \ref{fig:VRinterface}B). The annotations included in the dataset are automatically loaded in and are encoded on a color map with proximity triggered labels appearing at each annotation's average position to avoid clutter (Figure \ref{fig:VRinterface}C). 

Corvo has two data viewing modes: one for metadata and one for gene expression, accessed through a controller binding. Gene expression is encoded on a color scale and normalised to maintain contrast in weakly expressed genes. Metadata annotations are also encoded as color and can be toggled through sequentially. 
\\\\
In gene expression mode, highly differentially expressed genes can be found in three ways:
\begin{enumerate}
    \item The 10 genes with the lowest p-value for each category in each annotation can be pre-computed from the launcher interface prior to launch and saved for future use (Figure \ref{fig:LauncherInterface}A). Upon selecting an annotation label with the controller, the precomputed most differentially expressed genes, and their associated p-values and log fold changes, are displayed to the user interactively, attached to the controller as a clipboard (Figure \ref{fig:VRinterface}A),

    \item Individual genes can be searched for using voice input (\dg{3}).\footnote{In order to address potential privacy concerns, we have opted to use the Vosk offline speech-to-text engine, such that no voice input or recognised words or sentences are leaving the users computer. See \href{https://alphacephei.com/vosk/}{alphacephei.com/vosk/} for details.}.The 5 results with the highest recognition confidence are then parsed for a matching gene, which when found is added to a gene group that can be loaded into the scene from the menu attached to the left controller (Figure 1A, 2), and finally

    \item Cells can be manually selected using the lasso or selection sphere (Figure \ref{fig:VRinterface}C), where the intersection of the tools with the cells in the dataset adds them to the current selection. The selected and unselected cell sets can then be compared using the Welch t-test, that finds the 10 most differentially expressed genes in the set.
\end{enumerate}
All selections can be reset by selecting \emph{reset} on the left controller. Further sub-annotations that may be present are presented in a key in the field of view (Figure \ref{fig:VRinterface}D), and as proximity-triggered labels at the average cluster position. The current gene set can be toggled through, with the current gene’s name and unnormalized range displayed on a scale in the field of view of the user (see Figure \ref{fig:VRinterface}B).

\section{Summary and Future Work}
In this paper, we have introduced \emph{Corvo}, a VR visualization and analysis tool for dimensionally-reduced CellxGene single-cell datasets. Compared to existing VR solutions, Corvo provides no-code dataset ingestion, multi-modal user input, and is interoperable with the existing Python-based data science ecosystem.

While Corvo is focused on CellxGene-hosted datasets at the moment, both dataset loading and preprocessing are designed in a modular way. Corvo can therefore be easily extended make use of other data sources, such as other AnnData \texttt{h5ad} files. Already, Corvo will try to process any \texttt{h5ad} datasets drag-and-dropped into the window by the user, even if they are not a part of CellxGene. Furthermore, while Corvo already implements a large part of the feature set of the web-based CellxGene interface, but we have not yet reached feature parity. In the future, we would therefore like to add metadata histograms, store gene sets persistently for later re-use, enable customization of the embedding used, and add the option of running the Welch t-test on different subsets of cells.

\section*{Software Availability}
The software is available as source distribution from Github. The core is available at \href{https://github.com/scenerygraphics/corvo-core}{github.com/scenerygraphics/corvo-core}, and the Python front-end at \href{https://github.com/scenerygraphics/corvo}{github.com/scenerygraphics/corvo}. Both projects are available under the open-source BSD license, and we highly welcome contributions to the projects.

\section*{Acknowledgements}
LH would like to thank the Quake, Royer, and Sbalzarini labs for hosting and mentoring during much of the work on Corvo. LH would like to thank the CZ Biohub and the Center for Systems Biology Dresden for funding. LH also thanks Angela Pisco for guidance on accurate representation and processing of single-cell data. This work was partially funded by the Center for Advanced Systems Understanding (CASUS), financed by Germany’s Federal Ministry of Education and Research (BMBF) and by the Saxon Ministry for Science, Culture and Tourism (SMWK) with tax funds on the basis of the budget approved by the Saxon State Parliament. 

{\footnotesize
   \bibliographystyle{abbrv-doi-hyperref-narrow}
   \bibliography{refs}
}

\end{document}